\documentstyle[prl,twocolumn,aps]{revtex}
\begin{document}
\draft
\title{Scaling Analysis of Magnetic Field Tuned Phase Transitions in
One-Dimensional Josephson Junction Arrays}
\author{Watson Kuo$^{1,2}$ and C. D. Chen$^{2}$\cite{email}}
\address{$^1$Department of Physics, National Tsing-Hua University,
Hsin-Chu, 300, Taiwan, ROC \\ $^2$Institute of Physics, Academia Sinica,
Nankang, 115, Taipei, Taiwan, ROC}
\date{March 12, 2001}
\maketitle
\begin{abstract}
We have studied experimentally the magnetic field-induced
superconductor-insulator quantum phase transition in one-dimensional
arrays of small Josephson junctions. The zero bias resistance was found to
display a drastic change upon application of a small magnetic field; this
result was analyzed in context of the superfluid-insulator transition in
one dimension. A scaling analysis suggests a power law dependence of the
correlation length instead of an exponential one. The dynamical exponents
$z$ were determined to be close to 1, and the correlation length critical
exponents were also found to be about 0.3 and 0.6 in the two groups of
measured samples.
\end{abstract}


A 1D array of small Josephson junctions (JJ) provides an ideal testing
ground for the $T$=0 quantum phase transition\cite{chow,oude}.
Theoretically, a $d$-dimensional JJ array can be mapped to the classical
($d$+1)-dimensional $XY$ model, however, the types of governing phase
transitions may not be the same in different dimensions\cite{brad,sondhi}.
In the 2D $XY$ model, for instance, the exponentially dependent
correlation length of Kosterlitz-Thouless-Berezinskii(KTB) transition
should lead to scaling properties different from those in a 3D
model\cite{KT}. While scaling properties in 2D JJ arrays have been shown
to exhibit an correlation length which corresponds to an underlying power
law\cite{zant,ccd}, no 1D arrays experiment has been reported to date. In
this study we investigate the scaling behavior in 1D JJ arrays, and
report a power law dependent correlation length. In addition, correlation
length exponents $\nu$ are determined at approximately 0.3 and 0.6.

As shown in Fig.\ \ref{a2iv}(a), the measured arrays were composed of
I-shaped aluminum islands, whose sizes as defined by electron beam
lithography were on the order of 1 $\mu $m. Each Al island has two tunnel
junctions connected in parallel to its nearest neighbors, and forms a
SQUID which is referred to as a unit cell. Being fabricated on the same
chip, each group of measured arrays (denoted as group A and B) having
different cell numbers, $N$, should have almost similar controlled
parameters. Thus, length dependence of the phase transition can be
investigated. The normal state resistances $R_{T}$ of each cell for the
arrays, as listed in Table 1, are determined at high bias,
$V>N(2\Delta^{0}/e)$. The resistances for arrays in a group are quite
similar, confirming the uniformity of the fabricated arrays.

When magnetic field $B$ is applied perpendicularly to SQUID loops with
area $A$, the Josephson coupling energy $E_{J}$ can be tuned periodically
as $E_{J}=E_{J}^{0}\cos (\pi AB/\Phi _{0})$. The zero field Josephson
coupling $E_{J}^{0}$ is determined using the $T=0$ Ambegaokar-Baratoff
formula $E_{J}^{0}=(\Delta ^{0}/2)(R_{Q}/R_{T})$ with a superconducting
gap $\Delta ^{0}$ of about 200$\mu $eV, and a resistance quantum $R_{Q}$
of $h/4e^{2}\approx 6.5$k$\Omega$. Accordingly, the $E_{J}^{0}$ values are
approximately 130$\mu$eV for arrays in group A, and 650$\mu$eV for arrays
in group B. From the SEM image one can estimate the junction area, and
infer a junction capacitance $C$ of 3.0 $\pm $0.8fF by using a specific
capacitance of 45 fF/$\mu $m$^{2}$\cite{delsing}. With this capacitance,
the charging energy for Cooper pairs, $E_{CP}\equiv (2e)^{2}/2C$, is about
106$\pm $35$\mu $eV, giving $E_J^0/E_{CP}$ values of about 6 and 1.3 for
arrays in groups A and B, respectively.

The transport measurements were conducted in a dilution refrigerator at
temperatures ranging between 40mK and 1.5K.  Samples were mounted inside
an electromagnetic-field-sealed compartment within an oxygen-free copper
holder. Any external noise was carefully prevented from reaching the
sample compartment. The signal leads were filtered by $\pi$-section
low-pass feedthrough filters, which were on the top of the cryostat and
also with ThermoCoax cables from room temperature to mK temperature of
sample compartment. To minimize the line frequency noise, a
battery-powered preamplifier mounted on the top of cryostat was used.
Furthermore, taking advantage of the common mode noise rejection, the
entire measurement circuit was placed symmetrically with respect to its
ground. Zero bias resistance $R_{0}$ was taken from the slope of $I(V)$
characteristics at a very small bias, and it was further confirmed by
using a lock-in technique at a frequency of 1.7Hz with an excitation of
about 20nV. Sweeping a wide range of magnetic fields at 40mK, we
determined the periodicity of magnetoresistance oscillation to be 9.15
Gauss. With this period $\Delta B$, we denote the field as a dimensionless
filling number $f=B/\Delta B$, which represents an average number of flux
quantum in one cell. At integer values of $f$, the arrays are most
conductive with $R_{0}$ at a minimum, while at half integer $f$-values,
the arrays become most resistive with $R_{0}$ at a maximum.

Figure\ \ref{a2iv}(b) shows $I(V)$ characteristics for array A1 measured
at $f=0\sim$0.5. For array A1 at $f=0$ ({\it i.e.} the most
superconducting curve), deviations of the supercurrent of consisting
junctions are quite small, reaffirming the uniformity of these arrays.
However, even at $f$=0, the array is not truly superconducting, but has a
finite zero bias resistance of about 0.9 k$\Omega$.  The supercurrent
decreases with $f$ and diminishes at $f\simeq $0.45, above which the
supercurrent-type structure turns into a Coulomb-blockade-type structure,
with zero bias resistance reaching a maximum value of about 17 M$\Omega$
at $f$=0.5. The evolution from one structure to the other is best
represented by Fig.\ \ref{a2iv}(c), which shows a smooth crossover from
dip to hump structure with differential resistance $R_{d}\equiv dV/dI$ as
a function of bias voltage, and with a flat $R_{d}(V)$ curve separating
the two limits.

Due to a smaller $E_J^0/E_{CP}$ value, the $I(V)$ characteristics for
array B1, depicted in Fig.\ \ref{b1iv}(a), have a higher $R_{0}$ of about
100 k$\Omega$ at $f$=0 and 50 G$\Omega$ at $f$=0.5. The evolution of
$I(V)$ characteristics, from the supercurrent-type structure to
Coulomb-blockade-type structure, for this array is quite different than
that of array A1:  As $f$ increases beyond $f$=0.20, a small Coulomb gap
in the $I(V)$ characteristic appears in the begining, signifying
competition between Josephson coupling and Coulomb blockade of Cooper pair
tunneling. This behavior can be clearly seen in Fig.\ \ref{b1iv}(b), which
shows, in addition to the simple dip and hump structures (as in Fig.\
\ref{a2iv}(c)), a coexistence of the dip and hump structures in the
$R_{d}(V)$ curves. From these $R_{d}(V)$ curves, supercurrent and Coulomb
blockade thresholds are plotted as a function of the filling numbers and
are shown in Fig.\ \ref{b1iv}(c).  Note that the two curves cross each
other, a feature different from that reported in Ref.\cite{chow} with a
considerably larger $E_J^0/E_{CP}$ value (about 6.1) than those of our
arrays in group B.

The temperature dependences of zero bias resistance at various filling
numbers for the arrays A2 and B1 are depicted in Fig.\ \ref{RT}. For the
most insulating case, the conductance fits standard Arrhenius form
between 1K and 150mK with an energy barrier of about 120$\mu $eV, which
is very close to $E_{CP}$. This suggests that the dynamics is dominated by
simple thermal activation of Cooper pairs, because the strength of the
Josephson coupling is suppressed to the minimum. At even lower
temperatures, shorter arrays show a saturation of resistance probably due
to the finite size effect. For the most superconducting case ({\it i.e}.
$f$=0), the resistance decreases with decreasing temperature and levels
off at $T<$0.5K. At low temperatures with $f\neq$0, $R_0(T)$ increases
even with decreasing temperature.  A close inspection reveals a separation
point $R_0(T\rightarrow0)\simeq R_Q$, above which $R_0(T)$ curves go
upward with decreasing temperature (leveling off for the case of shorter
arrays) and below which $R_0(T)$ curves simply level off. Similar results
are reported by Haviland {\it et al.} \cite{chow}. We emphasize that this
leveling off behavior is not due to any unwanted high-frequency noises in
the measurement system for these reasons: 1. the leveling-off temperatures
for the S and I sides are not the same, and 2. the leveling-off
temperatures for the arrays in the two groups are not the same.  This
leveling-off behavior cannot be accounted for by heating effects because
the measured resistance is almost the same for both AC and DC
measurements. There are possible reasons that may bring in a finite
leveling-off zero bias resistance: the finite size effect, the vortex
macroscopic quantum tunneling (MQT)\cite{gio}, and the charge Coulomb
interaction\cite{duan}. The finite size effect can be ruled out because
the leveling-off temperature does not change in arrays with different
lengths. Even if it does involve, it should set in at the same temperature
for both S and I sides, contrary to the $R_0(T)$ curves. The vortex MQT,
observed in wire experiments \cite{gio}, has a pronounced effect in 1D
systems and can result in broadening of superconductor
transition. However, MQT
should be exponentially suppressed by dissipation, which is quantified as
a ratio between the quantum resistance and the junction tunneling
resistance, $R_Q/R_T$. Arrays in group A have a dissipation about five
times larger than that of arrays in group B, and should have a diminishing
vortex MQT. The fact that a higher leveling off temperature for arrays in
group A suggests that MQT is not responsible for this behavior at this
temperature.

The last effect seems to be a reasonable one if the charge soliton picture
is taken into account. While the origin model\cite{sondhi} describing
phase transitions in 1D Josephson junction arrays considered only
an``on-site" Coulomb interaction, in reality, the long range interaction
due to a nonzero island-to-ground capacitance $C_0$ cannot be ignored. For
finite $C/C_0$, it has been shown \cite{likh} that the charge will spread
out and extend to a characteristic charge soliton length of
$\Lambda=\sqrt{C/C_0}$.  The result of the long range Coulomb interaction
may lead to finite array resistances at low temperatures, driving even the
superconducting phase to a metallic one\cite{duan}.  To investigate the
effect of this charge soliton picture, we made nearby grounded electrodes
along each array to control $C_0$.  The distance between electrodes and
arrays for samples in group A is 8 times longer than that for arrays in
group B, leading to a smaller $C_0$ and thus a longer soliton length. This
is qualitatively consistent with our experimental finding that arrays in
group A have higher leveling off temperatures than that of group B.

For $T>0.5$K, the sign of $dR_{0}/dT$ changes from positive to negative
value depending upon increasing filling numbers, signifying that the
system undergoes a quantum superconductor-insulator transition. According
to the theory of superfluid-insulator transition in 1D systems \cite{cha},
right at critical point $f=f^{\ast }$, the resistance is linearly
proportional to the temperature. Experimentally, $f^{\ast }$ is identified
as the filling number where the extrapolation of $R_{0}(T)$ curve passes
$R_{0}=0$, $T=0$ point. As a supplementary clue, we notice that at the
base temperature, this critical filling number corresponds to an onset of
Coulomb blockade threshold voltage (see Fig.\ \ref{b1iv}(c)). We thus
interpret these phenomena as evidence of a magnetic field-tuned SI phase
transition with a critical filling number $f^{\ast}$ in our measured
arrays.

In a non-interacting model, the Hamiltonian of a 1D array of small JJ can
be mapped to a classical 2D $XY$ model. Theoretically, the dimensionless
coupling constant $K$, playing the role of the temperature
in the classical model, is related to the
charging energy $E_{CP}$ and Josephson coupling energy $E_{J}$ as,
$K=\sqrt{E_{J}/2E_{CP}}$ in the quantum system\cite{sondhi}. The 2D $XY$
model has a KTB type
transition. Below the transition temperature $T_{{\rm KTB}}$, the spins
form vortex-antivortex pairs, while above $T_{{\rm KTB}}$, the pairs
dissociate and the whole system becomes a vortex plasma\cite{KT}. Note
that the topological spin vortex in the 2D $XY$ model represents the phase
slip in 1D JJ arrays. In our system, the corresponding KTB transition
takes place at critical coupling energy $E_{J}^{\ast }$, which is achieved
by the tuning of external magnetic field. In the region $E_{J}<E_{J}^{\ast
}$, corresponding to the pair-dissociation phase at $T>T_{{\rm KTB}}$, the
long range order in phase vanishes, and phase fluctuations render
insulating 1D JJ arrays. According to the model, the transition takes
place at $E_{J}^{\ast }/E_{CP}=8/\pi ^{2}\simeq 0.81$ \cite{brad}; however
the experimental values are slightly larger than the theoretical one.
Despite the scattering of $E_{J}^{\ast }/E_{CP}$ values in group A, the
values in group B are well consistent to one other. The difference between
the two groups can be explained with the effect of dissipation. At
different strength of dissipation, it is expected that the critical point
of SI phase transition will also be different \cite{korsh}.

The finite-temperature scaling law of quantum phase transitions asserts
that ${\cal O}(k,\omega ,K,L_{T})=L^{d_{{\cal O}}/z}{\cal
O}(kL_{T}^{1/z},\omega L_{T},L_{T}/\xi _{T}),$ where $L_T=\hbar \beta$ is
a finite length on the imaginary time axis. Some of the terms can be
neglected: the wave vector $k$ is assumed zero, the scaling dimension
$d_{{\cal O}}=2-d=1$ \cite{cha}, and $\omega=0$ in DC measurements. We
thus obtain a concise finite-size scaling form for zero bias resistance,
$R_0(f,T)=T^{1/z}\widetilde{R_0}(1/T\xi^z)$, where $\xi$ is a function of
$f$. To examine the correlation length dependence, we rewrite the scaling
form as $R_0(\delta,T)=T^{1/z}\widetilde{R_0}(T_1/T)$ where
$\delta=(K-K^{\ast})/K^{\ast}\sim (f-f^{\ast })$ is the distance from the
transition point and $T_1$ are field dependent scaling parameters (see
\cite{Analysis} for details). In this way, one obtains a clear power law
dependence of $T_1$ on $\delta$ over one order of magnitude, suggesting a
power law dependence of the correlation length. This is in contradiction
to the exponential dependence predicted in KTB transition and other
theories\cite{fisher}.

Based on this analysis, the correlation length $\xi$ can be written as
$\xi=|\delta|^{-\nu}$ where $\nu$ is a critical exponent. Finally, the
scaling law can be written with a scaling function $\widetilde{R_0}$ as

\begin{equation}
R_0(f,T)=T^{1/z}\times \widetilde{R_0} \left(\frac{\delta}{T^{1/\nu
z}}\right). \label{RTSeq}
\end{equation}

The exponent $\nu z$ is determined from the slope of the power law
fitting. For the seven arrays measured, we find that the $\nu z$ are
different varying from 0.30 to 0.60 for different arrays. Nevertheless,
the same values of $\nu z$ are obtained from both S and I sides. Assuming
a $z$ value of unity and using Eq.\ (\ref{RTSeq}) with $\nu z$ and
$f^{\ast}$ obtained from above, the scaling curves shown in Fig.\
\ref{RTS} can be obtained. In addition, the scaling function form is found
to be $\widetilde{R_0}(x)\propto e^{\alpha x}$, with $\alpha\simeq\pm$10
and $\pm$2.5 (in unit of ${\rm K}^{1/\nu z}$, `+' for I-phase and `$-$'
for S-phase), for arrays in groups A and B, respectively. This form for
resistance is similar to results deduced from variable-range hopping(VRH)
for the Bose glass\cite{fisher}. However, the VRH mechanism should not be
accounted for our system since the exponent on $T$ is larger than 1. The
fact that the scaling function in one phase is symmetrical to that in the
other phases suggests that the S phase and the I phase plays a dual role
at zero bias. To refine the critical filling number and the scaling
exponents, one begins with a trial scaling form, namely,
$R_0(f,T)=AT^{1/z}\exp ( \kappa(f)/T^{1/\nu z} ) $. By noting that $A$ is
$f$-independent constant and that $\kappa$ is zero at $f^{\ast}$, the $\nu
z$ and $f^{\ast}$ values can be unambiguously determined. The only
adjustable parameter, $z$, can then be determined from the scaling curves.
This parameter can be determined to a reasonable accuracy; a smaller $z$
would give better scaling on the S-side, whereas a larger $z$ would result
in better scaling on the I-side.  This fine-tuning procedure slightly
modifies $\nu z$ and $f^{\ast}$ values, and gives a $z$ value of about
0.85$\pm$0.05 for array B1.

To summarize, we observed a magnetic field-tuned SI phase transition in
the system of 1D small Josephson junction arrays, and have, for the first
time, made scaling analysis on such a system. Near the critical point, the
$R_0(T)$ scaling analysis indicates a power law dependent correlation
length. The exponents $\nu z$ are found to be 0.3 to 0.6 with $z$ close to
1, implying an isotropy in spatial and time dimensions. The value of
correlation length exponent $\nu$ contradicts what expected under current
theory of 1D Boson system, $\nu\geq 2/d$ and $\nu=\infty$\cite{fisher}.
The regions of lower temperature, where probably undermined complicated
physics to make scaling fail, is away from the scope of the
non-interacting model. These results suggest that certain important
physics has not been unearthed in this system.

Fruitful discussions with N. Trivedi, L. Balents, A. Schakel, P. Phillips,
S. K. Yip, C. M. Ho, and T. K. Lee are gratefully acknowledged.  We have
utilized facility at the National Nano Device Laboratories.  This research
was partly funded by the Nation Science Council No. 89-2112-M-001-033.

\begin{figure}[tbp]
\caption{(a)The SEM
image of an 1D JJ array. The overlapping areas between the `I' shaped
islands are the tunnel junctions. The scale bar at the bottom of the image
is 1$\protect\mu$m. (b) Evolution of $IV$ characteristics for arrays
A2, from superconducting behavior to insulating behavior, at selected
filling numbers between 0.0 and 0.5. (c) Dynamic resistance $R_d$ as a
function of bias voltage show a crossover from superconducting (bottom,
$f$=0.42) to insulating behavior (top, $f$=0.50)} \label{a2iv}
\end{figure}

\begin{figure}
\caption{$IV$
characteristics (a) and differential resistance $R_d(V)$ (b) for array
B1 for $f=0 \sim 0.5$ at 40mK. Notice the coexistence of the hump and dip
structures; this is not seen in arrays in group A ({\it c.f.} Fig. 1(c)).
The curves in (b) are shifted for clarity. (c) The $f$-dependence
of supercurrent (solid square) and threshold voltage (open diamond) for
array B1. Both are converted to the energy scale, {\it i.e.} ${\hbar
I_C}/2e$ and $eV$. The supercurrent is magnified 100 times.
}\label{b1iv}
\end{figure}

\begin{figure}[tbp]
\caption{The
$R_0(T)$ as a function of the filling number $f$ for the array A2(a) and
B1(b). At about 0.5K above, the array displays an $f$-tuned SI transition,
whereas at low temperatures the $R_0(T)$ curves in the S-side level off or
rise.} \label{RT}
\end{figure}

\begin{figure}[tbp]
\caption{$R_0(f,T)$ Scaling curve for array B1 for 0.5K$<T<$1K and
0$<f<$0.5. The inset shows the scaling curve for array A2.} \label{RTS}
\end{figure}

\begin{table}
\caption{Some important parameters of the measured arrays. The arrays that
are fabricated on the same chip and thus have similar junction parameters,
are categorized into one group. Note the closeness between the critical
filling numbers $f^*$ and the correlation length exponents $\nu$ for
arrays in a group.}
\begin{tabular}{|c|c|c|c|c|c|c|c|}
Sample & A1 & A2 & A3 & B1 & B2 & B3 & B4 \\ \hline
$N$&49&29&14&100&75&50&30\\ \hline $R_T$& 0.9&1.1&0.9&5.0&4.7&4.3&4.9\\
\hline $f^*$ &0.41&0.41&0.45&0.20&0.25&0.26&0.27\\ \hline $\nu
z$&0.3&0.3&0.3&0.6&0.6&0.5&0.5\\ \hline
$E_J^*/E_C$&2.0&1.6&1.1&1.05&0.97&1.02&0.88\\
\end{tabular}
\label{sample}
\end{table}


\begin{references}
\bibitem[\dag]{email} e-mail address: chiidong@phys.sinica.edu.tw

\bibitem{chow}  E. Chow, P. Delsing, and D. B. Haviland, Phys. Rev. Lett.
{\bf 81}, 204(1998);  D. B. Haviland, K. Andersson, and P. Agren,
cond-mat/0001143

\bibitem{oude}  A. van Oudenaarden and J. E. Mooij, Phys. Rev. Lett. {\bf 76}
, 4947(1996); A. van Oudenaarden, S. J. K. V\'{a}rdy, and J. E. Mooij, Phys.
Rev. Lett. {\bf 77}, 4257(1996)

\bibitem{brad}  R. M. Bradley and S. Doniach, Phys. Rev. B {\bf 30},
1138(1984);

\bibitem{sondhi}  S. L. Sondhi, S. M. Girvin, J. P. Carini, and D. Shahar,
Rev. Mod. Phys. {\bf 69}, 315(1997)

\bibitem{KT}  J. M. Kosterlitz and D. J. Thouless, J. Phys. C {\bf 6},
1181(1973); J. M. Kosterlitz, J. Phys. C {\bf 7}, 1046(1974)

\bibitem{zant}  H. S. J. van der Zant, F. C. Fritschy, W. J. Elion, L. J.
Geerligs, and J. E. Mooij, Phys. Rev. Lett. {\bf 69}, 2971(1992)

\bibitem{ccd}  C. D. Chen, P. Delsing, D. B. Haviland, Y. Harada, and T.
Claeson, Phys. Rev. B {\bf 51}, 15645(1995)


\bibitem{delsing}  P. Delsing, T. Claeson, K. K. Likharev, and L. S. Kuzmin,
Phys. Rev. B {\bf 42}, 7439(1990)

\bibitem{gio} N. Giordano, Phys. Rev. Lett. {\bf 61}, 2137 (1988)

\bibitem{duan} S. R. Renn and J.-M. Duan, Phys. Rev. Lett. {\bf 76}, 3400
(1996); B. J. Kim, G. S. Jeon, M.-S. Choi and M. Y. Choi, Phys. Rev. B
{\bf 58}, 14524 (1998)

\bibitem{likh} K. K. Likharev and K. A. Matsuoka, Appl. Phys. Lett. {\bf
67}, 3037 (1995)

\bibitem{korsh} S. E. Korshunov, Europhys. Lett. {\bf 9},
107(1989); W. Zwerger, Europhys. Lett. {\bf 9}, 421(1989)

\bibitem{cha}  M. P. A. Fisher, G. Grinstein, and S. M. Girvin, Phys. rev.
Lett. {\bf 64}, 587(1990); M.C. Cha, M. P. A. Fisher, S. M. Girvin, M.
Wallin, and A. P. Young, Phys. Rev. B {\bf 44}, 6883(1991)

\bibitem{fisher} M. P. A. Fisher, P. B. Weichman, G. Grinstein, and D. S.
Fisher, Phys. Rev. B {\bf 40}, 546(1989)

\bibitem{Analysis} The $T_1(\delta)$ dependence was determined in the
following way: We searched for an $f$-dependent variable $T_{1}$ which
could overlie the $R_0/T(T/T_{1})$ curves for various $f$ near $f^\ast$
into a single curve.

\end{references}
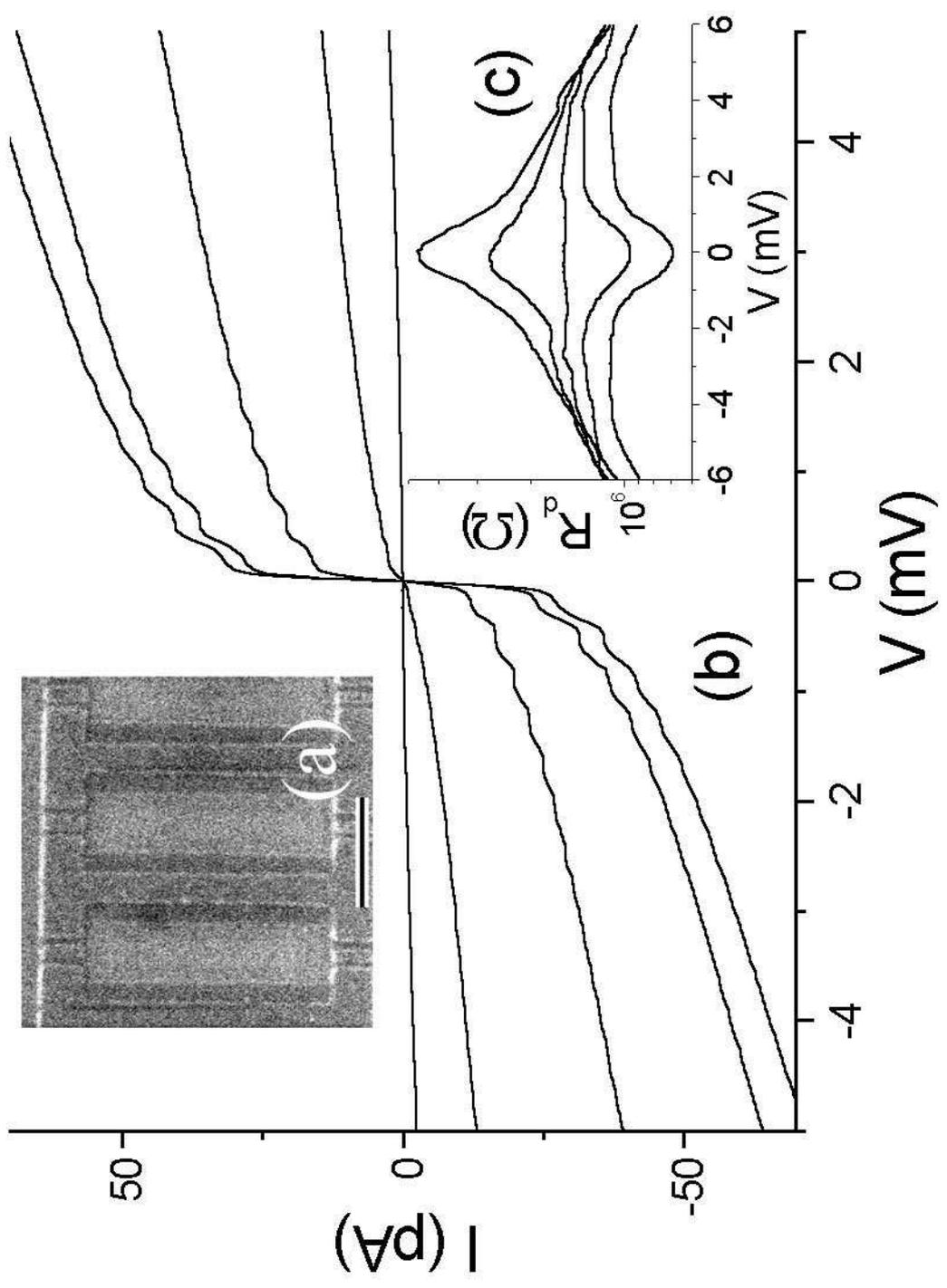
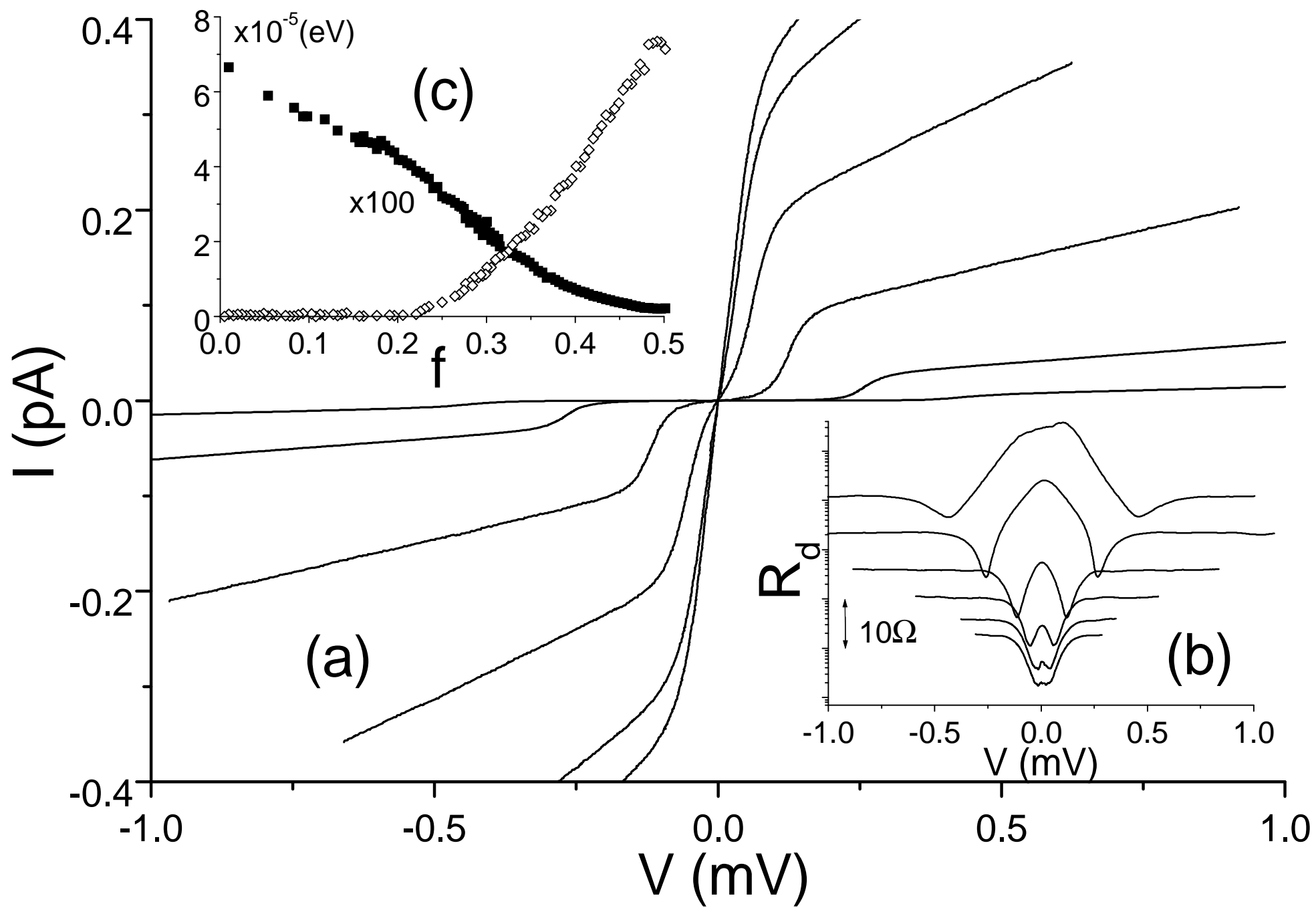
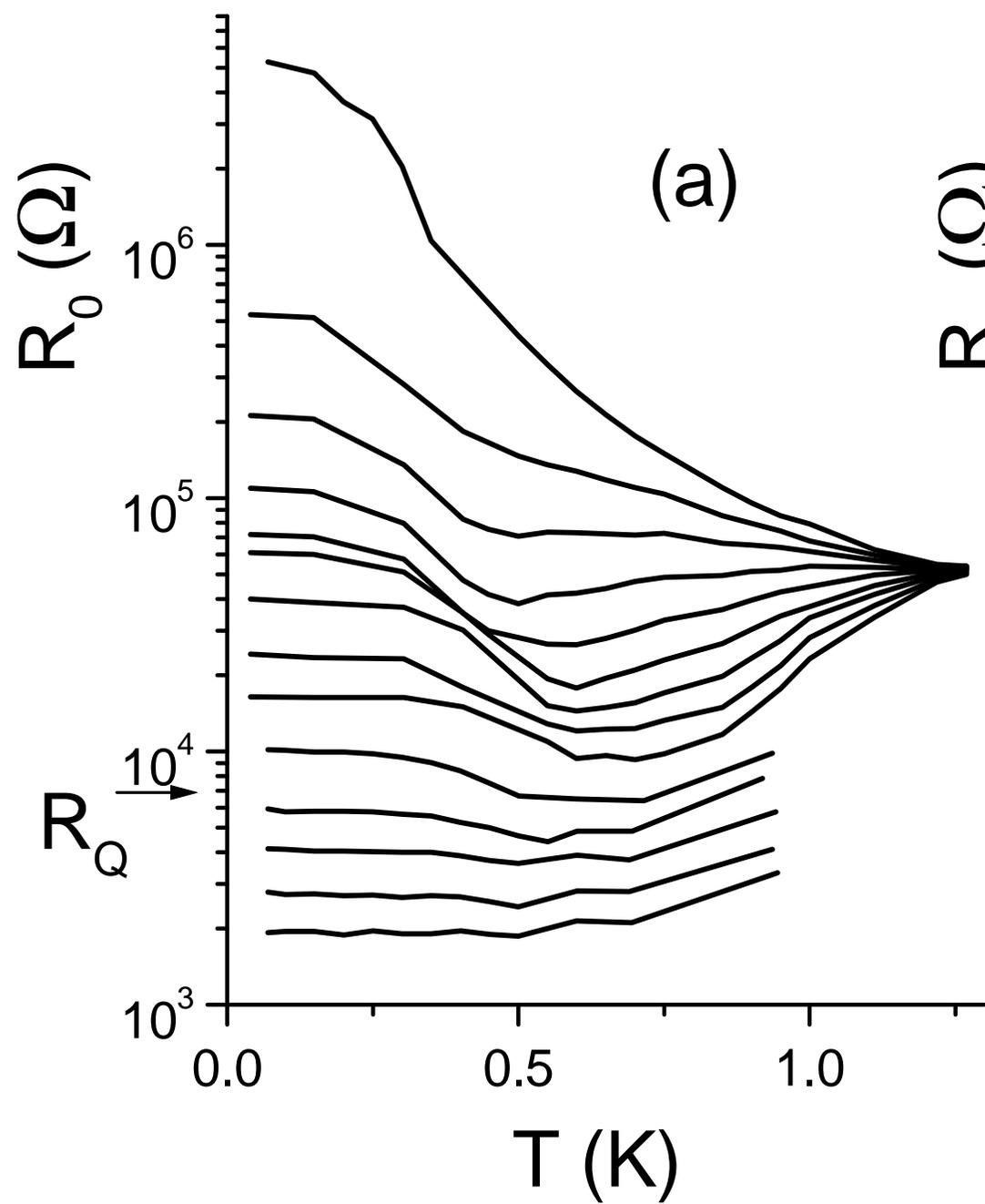
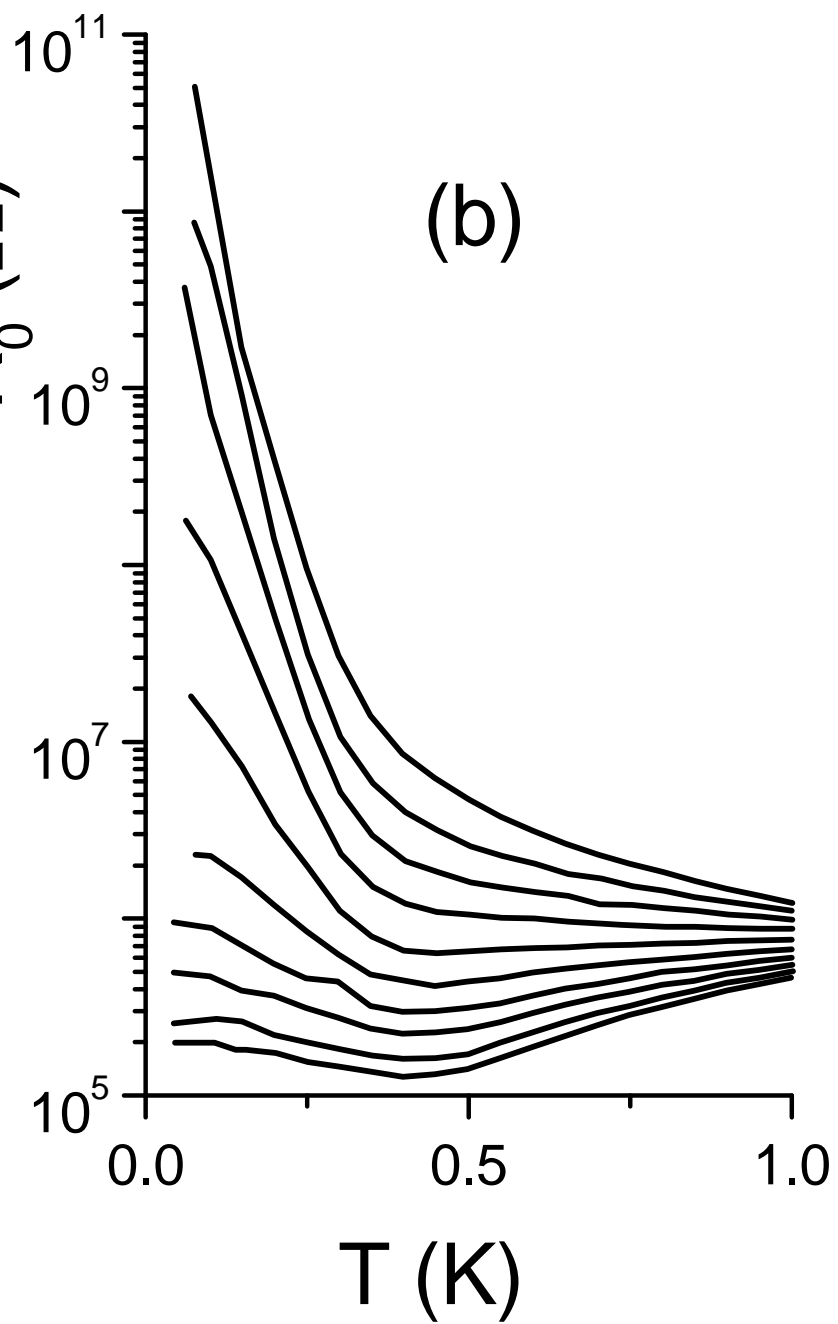
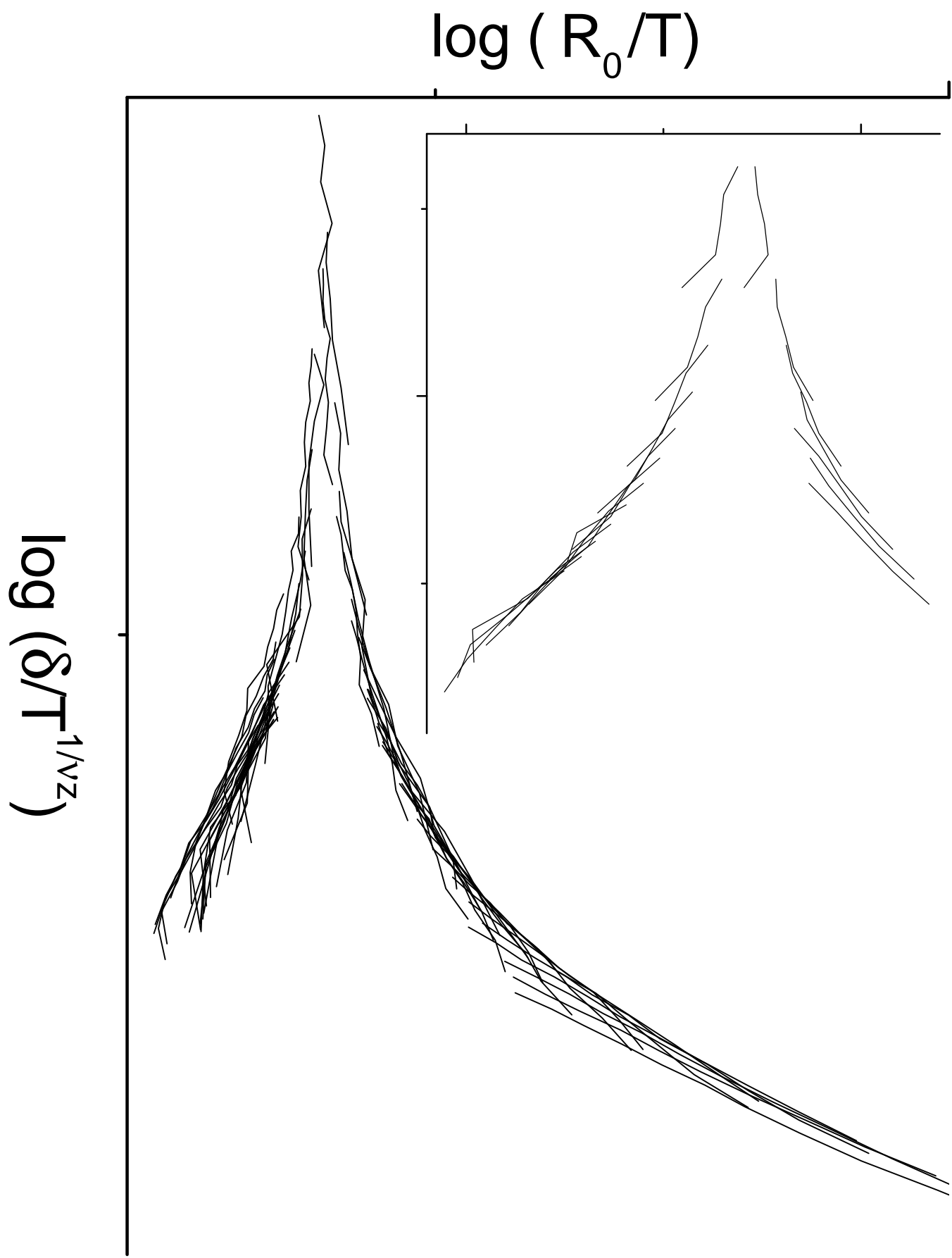
\end{document}